\documentclass[11pt]{article}
\usepackage{epsfig,here,hangcaption,rotating,fancyheadings,amssymb,amstex}
\input epsf
\catcode`\@=11
\def\marginnote#1{}

\def\draftlabel#1{{\@bsphack\if@filesw {\let\thepage\relax
  \xdef\@gtempa{\write\@auxout{\string
    \newlabel{#1}{{\@currentlabel}{\thepage}}}}}\@gtempa
    \if@nobreak \ifvmode\nobreak\fi\fi\fi\@esphack}
     \gdef\@eqnlabel{#1}}
\def\@eqnlabel{}
\def\@vacuum{}
\def\draftmarginnote#1{\marginpar{\raggedright\scriptsize\tt#1}}
\def\draft{\oddsidemargin -.5truein
        \def\@oddfoot{\sl preliminary draft \hfil
        \rm\thepage\hfil\sl\today\quad\militarytime}
        \let\@evenfoot\@oddfoot \overfullrule 3pt
        \let\label=\draftlabel
        \let\marginnote=\draftmarginnote

\def\@eqnnum{(\theequation)\rlap{\kern\marginparsep\tt\@eqnlabel}%
\global\let\@eqnlabel\@vacuum}  }
\def\preprint{\twocolumn\sloppy\flushbottom\parindent 1em
        \leftmargini 2em\leftmarginv .5em\leftmarginvi .5em
        \oddsidemargin -.5in    \evensidemargin -.5in
        \columnsep 15mm \footheight 0pt
        \textwidth 250mmin      \topmargin  -.4in
        \headheight 12pt \topskip .4in
        \textheight 175mm
        \footskip 0pt

\def\@oddhead{\thepage\hfil\addtocounter{page}{1}\thepage}
        \let\@evenhead\@oddhead \def\@oddfoot{} \def\@evenfoot{}
}
\def\titlepage{\@restonecolfalse\if@twocolumn\@restonecoltrue\o
necolumn
     \else \newpage \fi \thispagestyle{empty}\c@page\z@
        \def\thefootnote{\fnsymbol{footnote}} }
\def\endtitlepage{\if@restonecol\twocolumn \else  \fi
        \def\thefootnote{\arabic{footnote}}
        \setcounter{footnote}{0}}  
\catcode`@=12
\relax
\newcommand{\newc}{\newcommand}
\newc{\ETM}{E_T^{\mathrm{miss}}}
\newc{\MS}{M_{\mathrm{susy}}} 
\newc{\MSS}{M_{\mathrm{susy}}^{\mathrm{eff}}}
\newc{\MX}{M_{\chi}} 
\newc{\mest}{M_{\mathrm{est}}}
\newc{\sigsus}{\sigma_{\mathrm{susy}}}

\begin{document}
\topmargin-1.cm
%
\begin{titlepage}
\vspace*{-64pt}

\begin{flushright}
{SHEF-HEP/00-3,\\
June 2000}\\
\end{flushright}

\vspace{1.8cm}

\begin{center}

\large{\bf{Measuring the SUSY Mass Scale at the LHC}}\\
\vspace*{1.3cm}
\large{D.R. Tovey \footnote{e-mail: d.r.tovey@sheffield.ac.uk}}\\
\vspace*{0.4cm}
{\it\small Department of Physics and Astronomy, University of
Sheffield, \\ Hounsfield Road, Sheffield S3 7RH, UK.}\\

\end{center}

\bigskip
\begin{abstract}

An effective mass scale $\MSS$ for supersymmetric particles is defined
and techniques for its measurement at the LHC discussed. Monte Carlo
results show that, for jets + $\ETM$ events, a variable constructed
from the scalar sum of the transverse momenta of all reconstructed
jets together with $\ETM$ provides in most cases the most accurate
model independent measurement of $\MSS$ (intrinsic precision $\sim$
2.1 \% for mSUGRA models). The overall precision with which $\MSS$
could be measured after given periods of LHC running and for given
classes of SUSY models is calculated. The technique is extended to
measurements of the total SUSY particle production cross section
$\sigsus$.

\end{abstract}

\vspace{1cm}


\end{titlepage}

\setcounter{footnote}{0}
\setcounter{page}{0}
\newpage

\section{Introduction}

One of the principal motivations for construction of the Large Hadron
Collider is the search for low energy supersymmetry (SUSY)
\cite{nhk}. In a large class of models the interactions of SUSY
particles conserve R-parity, causing the Lightest Supersymmetric
Particle (LSP) to be neutral and stable. R-Parity conserving SUSY
events at hadron colliders are predicted to consist of cascade decays
of heavy, strongly interacting SUSY particles into lighter Standard
Model (SM) particles and two LSPs. This results in the classic
discovery signature of an excess of events containing jets, leptons
and large quantities of event missing transverse energy $\ETM$
\cite{hb}.

Should R-Parity conserving SUSY particles be discovered at the LHC the
next task would be to measure their properties. Importantly, such
measurements must be independent of the SUSY model and its parameters,
which are {\it a priori} unknown. This process is complicated by lack
of knowledge of the momenta of the two escaping LSPs in each event,
preventing direct reconstruction of SUSY particle masses. Consequently
other techniques are required which can measure masses or combinations
of masses indirectly. With sufficient integrated luminosity it should
be possible to look for edges in the invariant mass spectra of various
combinations of jets and leptons in SUSY events \cite{hb,fep}, but
initially the most effective technique is likely to be the use of
distributions of event transverse momentum $p_T$ and missing
transverse energy \cite{fep}. In this letter, the latter technique will
be investigated in detail and extended to SUSY models beyond the
minimal Supergravity (mSUGRA) models considered previously \cite{fep}.

\section{Measurement Technique}

Consider a heavy SUSY particle (mass $m_1$) produced in a
hadron-hadron collision. Assume further that this particle is boosted
along the beam-axis by the longitudinal momentum imbalance of the
event. If this particle undergoes a cascade decay to a lighter SUSY
particle (mass $m_2$) and a Standard Model particle (assumed
massless), then the transverse momentum $p_T$ of the SM particle in
the lab frame is related to $m_1$ and $m_2$:
\begin{equation}
\label{eq1}
p_T \propto \frac{1}{2} \left(m_1 - \frac{m_2^2}{m_1}\right).
\end{equation}
Variables based on the $p_T$ of SM particles in SUSY events are
therefore sensitive to SUSY particle masses, modulo smearing effects
arising from the true $\eta$ distribution of those particles.

Events in the jets + $\ETM$ + 0 leptons channel were used for this
study. The lepton veto requirement was imposed to reduce possible
systematics in the measurement arising from SM neutrino
production. Defining $p_{T(i)}$ as the transverse momentum of jet $i$
(arranged in descending order of $p_T$), the following four
measurement variables $\mest$ were studied:
\renewcommand{\labelenumi}{(\theenumi)}
\begin{enumerate}
 \item $\mest = |p_{T(1)}| + |p_{T(2)}| + |p_{T(3)}|
+ |p_{T(4)}| + \ETM$,
 \item $\mest = |p_{T(1)}| + |p_{T(2)}| + |p_{T(3)}| + |p_{T(4)}|$,
 \item$\mest = \sum_i |p_{T(i)}| + \ETM$,
 \item$\mest = \sum_i |p_{T(i)}|$.  
\end{enumerate}
The first variable is identical to the ``effective mass'' variable
($M_{\mathrm{eff}}$) defined in Ref.~\cite{fep}. The scalar sum of the
transverse momenta of only the four hardest jets was used due to the
predominantly four-jet nature of many SUSY events \cite{fep2}, while
the addition of the event $\ETM$ accounts for the $p_T$ carried away
by the LSPs.

Combining all jets in each detector hemisphere $j$ into one
pseudo-particle of transverse momentum $p_{T(j)}$ and invariant
mass $m_j$, a fifth variable was also defined:
\renewcommand{\labelenumi}{(\theenumi)}
\begin{enumerate}
\setcounter{enumi}{4}
 \item $\mest = \frac{1}{2} \sum_{j=1}^2 \sqrt{m_j^2 +
4p_{T(j)}^2 + 2\sqrt{(m_x^2+2p_{T(j)}^2).(m_j^2 + 2p_{T(j)}^2)}}$, 
\end{enumerate}
where $m_x = 100$ GeV/c$^2$. This variable approximates the mean of
the reconstructed masses of the two initial SUSY particles, assuming
that each moves close to the beam axis and decays into an LSP with
mass of order $m_x$.

\section{Definition of Mass Scale}

Due to the large number of different SUSY particles which can be
produced in any given event masses measured with these variables will
not correspond to those of any one particular SUSY state. Nevertheless
in many models the strongly interacting SUSY particles are
considerably heavier than states further down the decay chain. Hence
it is the decays of these particles which will contribute most to the
sum of $|p_T|$. For this reason we shall choose to define a SUSY
``mass scale'' $\MS$ as the weighted mean of the masses of the initial
SUSY particles, with the weighting provided by the production cross
section of each state:
\begin{equation}
\label{eq2}
\MS = \frac{\sum_i \sigma_i m_i}{\sum_i \sigma_i}.
\end{equation}
This definition differs from that used in Ref.~\cite{fep} but in the
limit where squarks or gluinos of a single mass dominate the
production cross section it gives the same result.

The above approximation breaks down for models where the lighter SUSY
particles are of similar mass to the strongly interacting states. We
shall attempt to compensate for this when using variables (1) - (4) by
defining an effective SUSY mass scale $\MSS$ in analogy with
Eqn.~(\ref{eq1}):
\begin{equation}
\label{eq3}
\MSS = \left(\MS - \frac{\MX^2}{\MS}\right),
\end{equation}
where $\MX$ is the mass of the LSP. For variable (5) the equivalent
expression is somewhat different:
\begin{equation}
\label{eq4}
\MSS = \sqrt{\MS^2 - \MX^2}.
\end{equation}  
With these definitions of the effective SUSY mass scale comparison of
experimental results with the predictions of a given SUSY model
requires knowledge of the model dependent particle mass spectrum and
production cross sections.

\section{Simulation and Event Selection}

Events were generated using PYTHIA 6.115 \cite{pyt} (SM background,
mSUGRA and MSSM signal) and ISAJET 7.44 \cite{isa} (GMSB
signal). Hadronized events were passed through a simple simulation of
a generic LHC detector. The calorimeter was assumed to have
granularity $\Delta \eta \times \Delta \phi = 0.1 \times 0.1$ over the
range $|\eta| < 5$, and energy resolutions $10\%/\sqrt{E} \oplus 1\%$
(ECAL), $50\%/\sqrt{E} \oplus 3\%$ (HCAL) and $100\%/\sqrt{E} \oplus
7\%$ (FCAL; $|\eta| > 3$). Jets were found with the GETJET \cite{isa}
fixed cone algorithm with cone radius $\Delta R = 0.5$ and $E_T^{cut}
= 50$ GeV.

Events in the jets + $\ETM$ channel were selected with the following
criteria:
\begin{itemize}

\item $\geq 4$ jets with $p_T \geq 50$ GeV

\item $\geq 2$ jets with $p_T \geq 100$ GeV

\item $\ETM \geq$ max(100 GeV,$0.25\sum_i p_{T(i)}$)

\item Transverse Sphericity $S_T \geq 0.2$

\item $\Delta \phi_{(\bf{p}_{\bf{T}(1)},\bf{p}_{\bf{T}(2)})} \leq
170^{\mathrm{o}}$

\item $\Delta \phi_{(\bf{p}_{\bf{T}(1)}+\bf{p}_{\bf{T}(2)},\bf{\ETM})}
\leq 90^{\mathrm{o}}$

\item No muons or isolated electrons with $p_T > 10$ GeV in $|\eta| <
2.5$.

\end{itemize}

Standard Model background events were generated for the following
processes: $\mathrm{t\bar{t}}$ ($5\times10^4$ events), W + jet
($5\times10^4$ events), Z + jet ($5\times10^4$ events) and QCD
2$\rightarrow$2 processes \cite{pyt} ($2.5\times10^6$ events). The
distributions of the $\mest$ variables for these events were then
compared with those for SUSY signal events generated from the mSUGRA
(minimal Supergravity \cite{nhk}), MSSM (Minimal Supersymmetric
Standard Model\footnote{A constrained version of the MSSM with 15 free
parameters (no additional D-terms, 3rd generation trilinear couplings
derived from masses) implemented in SPYTHIA \cite{spyt}}\cite{nhk})
and GMSB (Gauge Mediated SUSY Breaking \cite{dfs}) models. In each
case 100 points were randomly chosen from within the parameter space
of the model, and $1\times10^4$ events (a factor 10 greater than in
Ref.~\cite{fep}) generated for each point.

For mSUGRA models the region of parameter space sampled was identical
to that used in Ref.~\cite{fep}: 100 GeV $ < m_0 < $ 500 GeV, 100 GeV
$ < m_{1/2} < $ 500 GeV, -500 GeV $< A_0 <$ 500 GeV, 1.8 $<
\tan(\beta) <$ 12.0 and $ \mathrm{sign}(\mu) = \pm$ 1. For MSSM models
the choice of points was complicated by the requirement that models be
physically realistic (positive particle masses etc.) and be suitable
for this study (neutralino LSP). The masses of the strongly
interacting SUSY particles and sleptons were constrained to lie in the
range from 250 GeV/c$^2$ to 2000 GeV/c$^2$ while the mass parameters
of the partners of the electroweak gauge bosons were constrained to
lie in the range from 50 GeV/c$^2$ to the mass of the lightest
strongly interacting SUSY particle or slepton. $\tan(\beta)$ was
constrained to lie in the range 1.4 $< \tan(\beta) <$ 100.0. Finally,
given that the lightest SUSY particles in models with low $\MSS$ have
a high probability of discovery before LHC comes on line, only those
models with $\MSS > 250$ GeV/c$^2$ were used.

In mSUGRA and MSSM models the LSP is generally the lightest neutralino
$\chi^0_1$, but in GMSB models this role is taken by the gravitino. If
the Next-to-Lightest Supersymmetric Particle (NLSP) is neutral and
sufficiently long-lived to escape from the detector then the
phenomenology is similar to that for mSUGRA or MSSM models
\cite{gmsb}. If the NLSP is short-lived however then it decays to a
gravitino and the phenomenology is different. To test whether the
$\mest$ variables defined above are also sensitive to the effective
mass scales of these latter models points were chosen from within the
range of GMSB parameter space defined by: 10 TeV $ < \Lambda_m < $ 100
TeV, 100 TeV $ < M_m < $ 1000 TeV/c$^2$, 1 $ < N_5 < $ 5, 1.8 $<
\tan(\beta) <$ 12.0 and $ \mathrm{sign}(\mu) = \pm$ 1. The value of
$C_{grav}$, the ratio of the gravitino mass to that expected for only
one SUSY breaking scale \cite{isa}, was set to unity in all cases to
ensure rapid decays to the gravitino LSP. Again only those models with
$\MSS > 250$ GeV/c$^2$ were used.

In the mSUGRA and MSSM models a statistically significant excess of
signal events (S) above background (B) ($\sqrt{S+B} - \sqrt{B} \geq
5.0$ \cite{kras}) was found for the majority of points after the
delivery of only 10 fb$^{-1}$ of integrated luminosity (1 year of low
luminosity operation). In GMSB models the data indicate that greater
event statistics corresponding to at least 100 fb$^{-1}$ (1 year of
high luminosity running) would be required for discovery with these
particular cuts. It should be noted however that the use of this
channel and these cuts has been optimised for mSUGRA points. In GMSB
models with prompt decays to gravitino LSPs photon production (bino
NLSP) or lepton production (slepton NLSP) is common \cite{fep-priv}
and consequently many events were rejected by the lepton veto and jet
multiplicity requirements. In dedicated GMSB studies these
requirements should therefore be loosened in order to increase signal
acceptance. In this case measurement variables taking account of
lepton and/or photon $p_T$ should also be used to reduce systematic
measurement errors \cite{fep-priv}.

\section{Mass Scale Measurement}

The $\mest$ distributions of SUSY signal events in the models
considered here are roughly gaussian in shape (see Figs. (1) - (5) of
Ref.~\cite{fep}), in sharp contrast to the SM background which falls
rapidly with $\mest$. Fitting gaussian curves to the signal
distributions then provides estimates of their means which can be
compared with the effective mass scales $\MSS$ of the corresponding
SUSY models. The degree of correlation between the two variables gives
a measure of the intrinsic (systematic) precision provided by the
$\mest$ variable when measuring $\MSS$.

\begin{figure}
\begin{center}
\epsfig{file=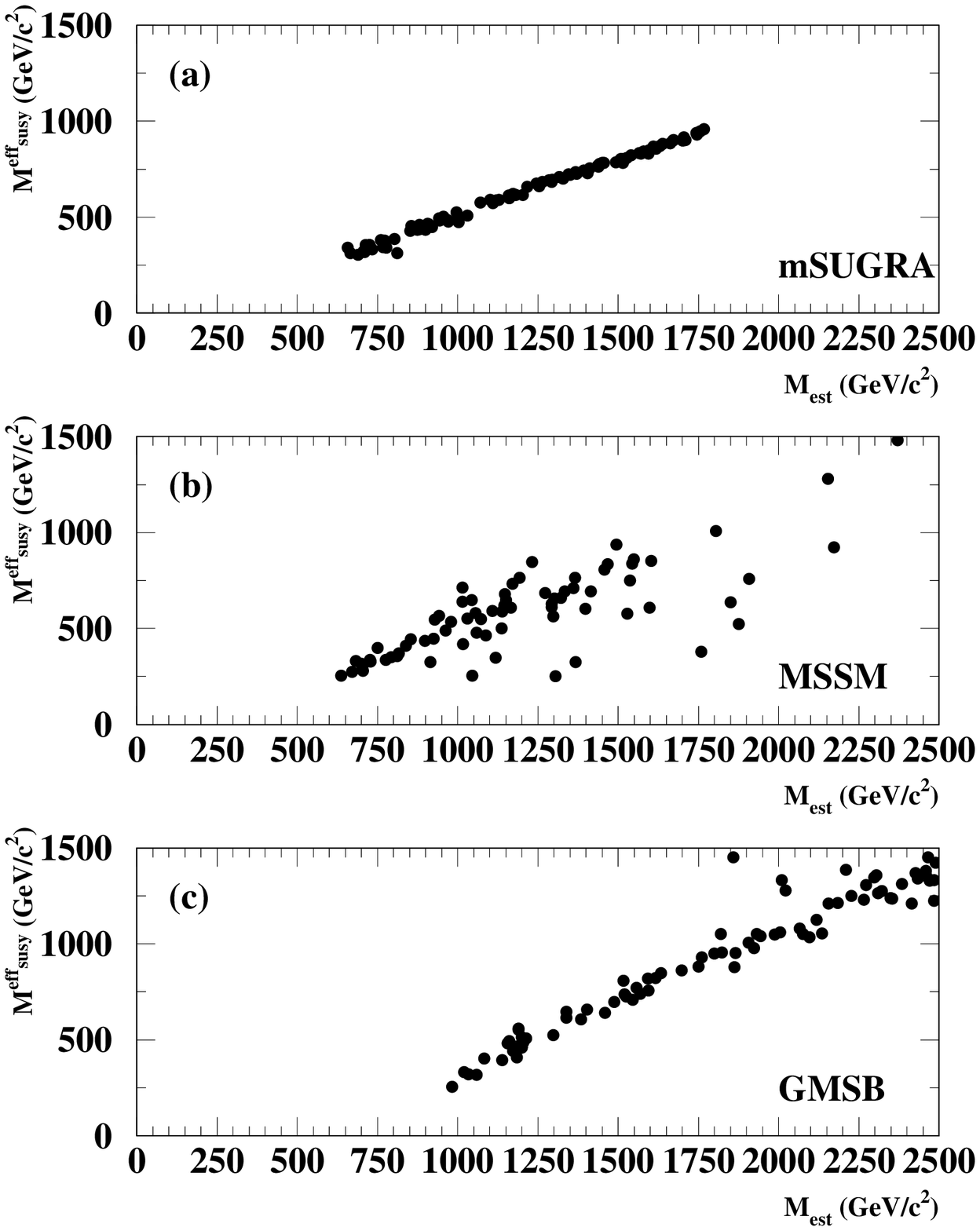,height=7.0in}
\caption{\label{fig1}}
\end{center}
\end{figure}

\begin{figure}
\begin{center}
\epsfig{file=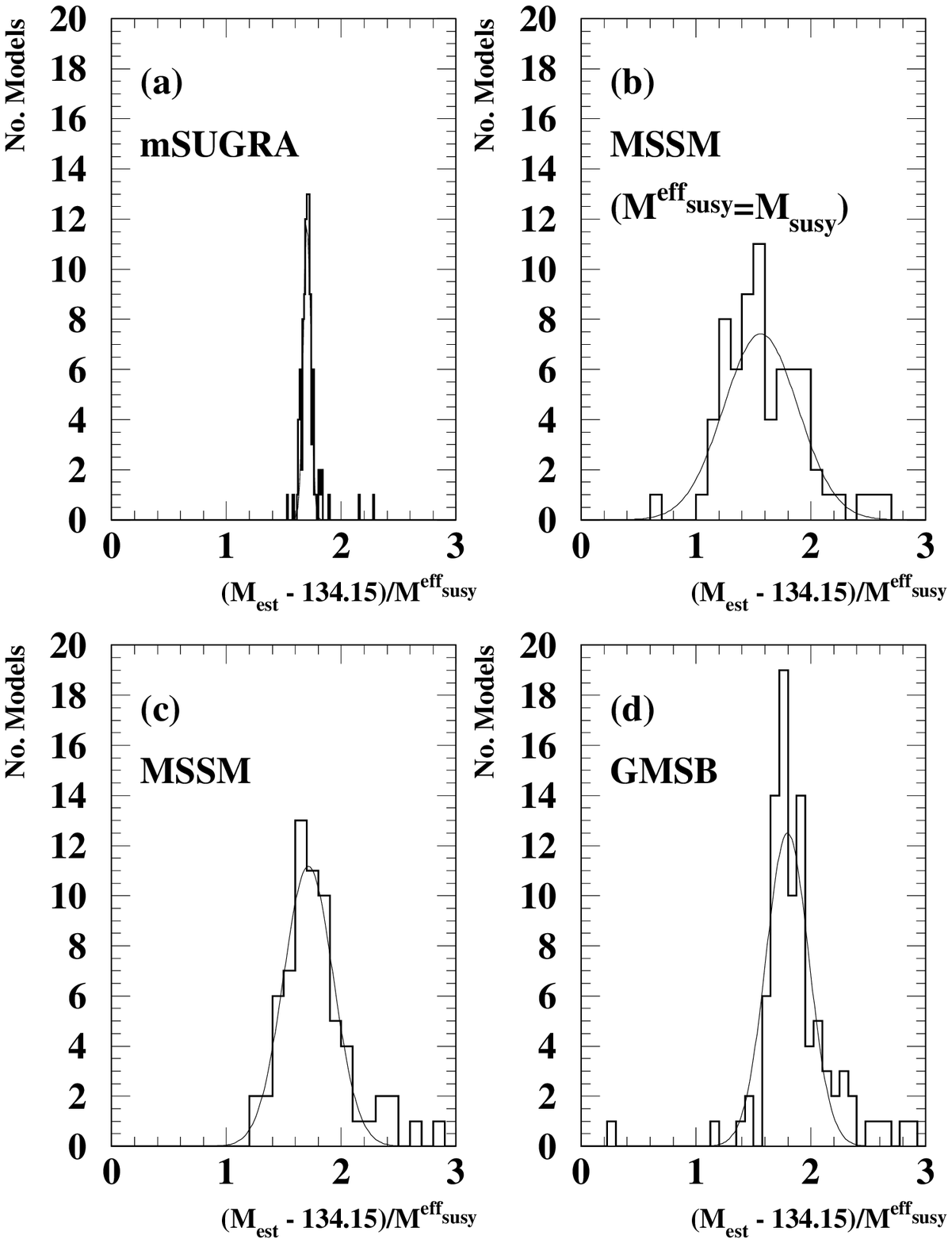,height=7.0in}
\caption{\label{fig2}}
\end{center}
\end{figure}

A typical scatter plot of $\MSS$ against $\mest$ (here variable (3))
for mSUGRA models is shown in Fig.~\ref{fig1}(a). The correlation
between the variables is clearly very good. To quantify the degree of
correlation a linear regression was performed on the data and the
points projected onto an axis perpendicular to the fitted
trendline. The distribution of the data along this line for mSUGRA
models is shown in Fig.~\ref{fig2}(a).

A scatter plot of $\MSS$ against $\mest$ variable (3) for MSSM models
is shown in Fig.~\ref{fig1}(b). In Figs.~\ref{fig2}(b) and (c) are
plotted correlation histograms derived from this figure using the
projection axis defined by the mSUGRA data. Fig.~\ref{fig2}(b) shows
the histogram obtained assuming $\MSS = \MS$, while Fig.~\ref{fig2}(c)
shows the equivalent histogram using the definition of $\MSS$ in
Eqn.~(\ref{eq3}). The smaller scatter in this latter case indicates an
improved measurement precision. An improvement is also obtained for
mSUGRA models. Here however it is smaller since $\MX$ is usually much
less than the masses of the strongly interacting SUSY particles
\cite{drees}. In GMSB models (Fig.~\ref{fig1}(c) and
Fig.~\ref{fig2}(d)) $\MX$ is negligible and so the two definitions of
$\MSS$ are always identical.

The correlation histograms were next fitted with gaussian
functions. The fitted values widths $\sigma$ were used to calculate
the intrinsic measurement precision by subtracting in quadrature the
width expected from finite event statistics (assumed to be given by
the rms of the errors on the fitted means of the $\mest$
distributions). Due to uncertainties in the expected statistical
scatter this correction could give unduly optimistic estimates of the
measurement precision if it were large. In all cases however the
corrections were found to be small ($\lesssim$ 33\%) and the effects
of such uncertainties neglected.

Intrinsic measurement precisions for $\MSS$ calculated using the
above technique for mSUGRA, MSSM and GMSB models and the five $\mest$
variables listed in Sec.~2 are presented in Table~\ref{tab1}. Variable
(3) ($\mest = \sum_i |p_{T(i)}| + \ETM$) provides the greatest
precision for mSUGRA and MSSM models, with the higher precision being
for mSUGRA models (2.1 \%). The poorer precision for MSSM models (12.8
\%) is due to their greater number of free parameters and hence the
smaller correlation between the particle masses. In the GMSB models
variable (4) ($\mest = \sum_i |p_{T(i)}|$) provides the greatest
measurement precision (6.1 \%), however variable (3) is reasonably
accurate (9.0 \%). This indicates that effective mass scale
measurements are also effective for models with gravitino LSPs.

It should be noted that for any given $\mest$ variable the fitted
means of the projected histograms (Table~\ref{tab1}) for mSUGRA, MSSM
and GMSB models are consistent to within the fitted widths. For this
reason it can be said that the variables provide {\it model
independent} measurements of the effective SUSY mass scale $\MSS$. The
expected measurement precision is SUSY model dependent (due to larger
widths for MSSM histograms than for mSUGRA and GMSB histograms) but
this is less troublesome when comparing measurements with theory
because in this case a particular SUSY model must be assumed.

With the intrinsic precision from Table~\ref{tab1} for $\mest$
variable (3) it is possible to estimate the overall (systematic +
statistical) precision for measuring effective SUSY mass scales in
mSUGRA, MSSM and GMSB models as a function of the mass scale and
integrated luminosity. Distributions of $\mest$ for signal +
background events were first constructed assuming integrated
luminosities of 10 fb$^{-1}$ (1 year low luminosity), 100 fb$^{-1}$ (1
year high lumi.)  and 1000 fb$^{-1}$ (10 years high lumi.). The mean
background distribution was then subtracted from each with an assumed
50\% systematic error \footnote{It is unlikely that the distribution
of background events (especially QCD) will be known with any certainty
from theory. Instead the distribution of low $\ETM$ events will likely
be measured and the results extrapolated into the high $\ETM$ region
\cite{cdf,d0}. 50\% is a conservative estimate of the systematic
uncertainty associated with this extrapolation.}. The resulting
distributions were again fitted with gaussian functions and the errors
on the fitted means added in quadrature to the intrinsic precision
calculated above and an estimated 1\% systematic error from
uncertainties in the measurement of the jet energy scale.

\begin{figure}
\begin{center}
\epsfig{file=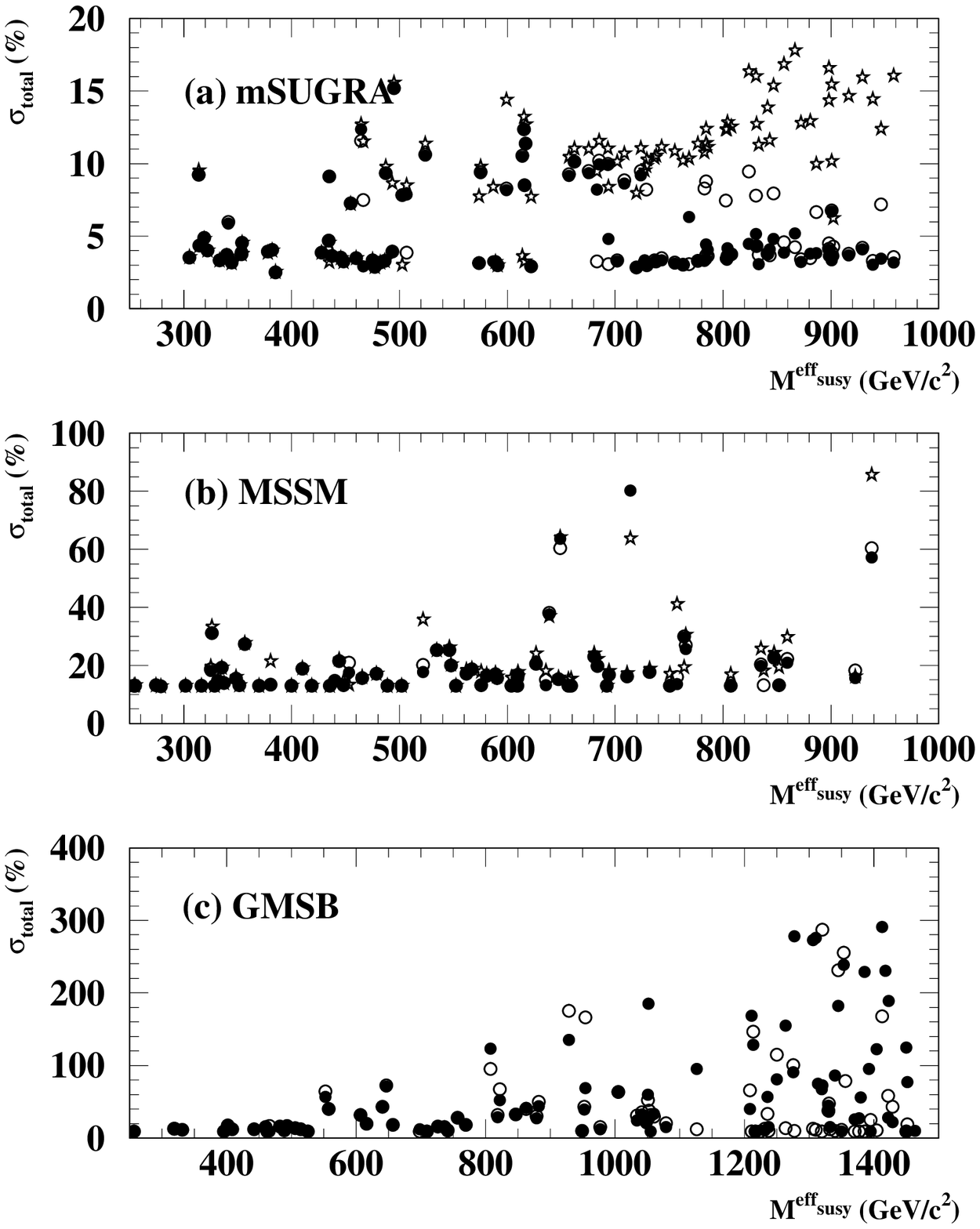,height=7.0in}
\caption{\label{fig3}}
\end{center}
\end{figure}

The results are plotted in Fig.~\ref{fig3} for mSUGRA, MSSM and GMSB
models. In the last case results for 100 fb$^{-1}$ and 1000 fb$^{-1}$
only are presented due to the poor statistical significance of GMSB
models in the jets + $\ETM$ channel at low integrated
luminosity. Precisions $\lesssim$ 15 \% (40 \%) should be achievable
in mSUGRA (MSSM) models after only one year of low luminosity running,
improving to $\lesssim$ 7 \% (20 \%) after one year of high luminosity
running. Due to the poor statistics obtainable from GMSB models,
particularly for high $\MSS$ values, measurement precisions $\lesssim$
50 \% are likely to be obtainable with these cuts only after the
delivery of 1000 fb$^{-1}$ of integrated luminosity and for $\MSS$
$\lesssim$ 1000 GeV/c$^2$.

\section{Cross Section Measurement}

\begin{figure}
\begin{center}
\epsfig{file=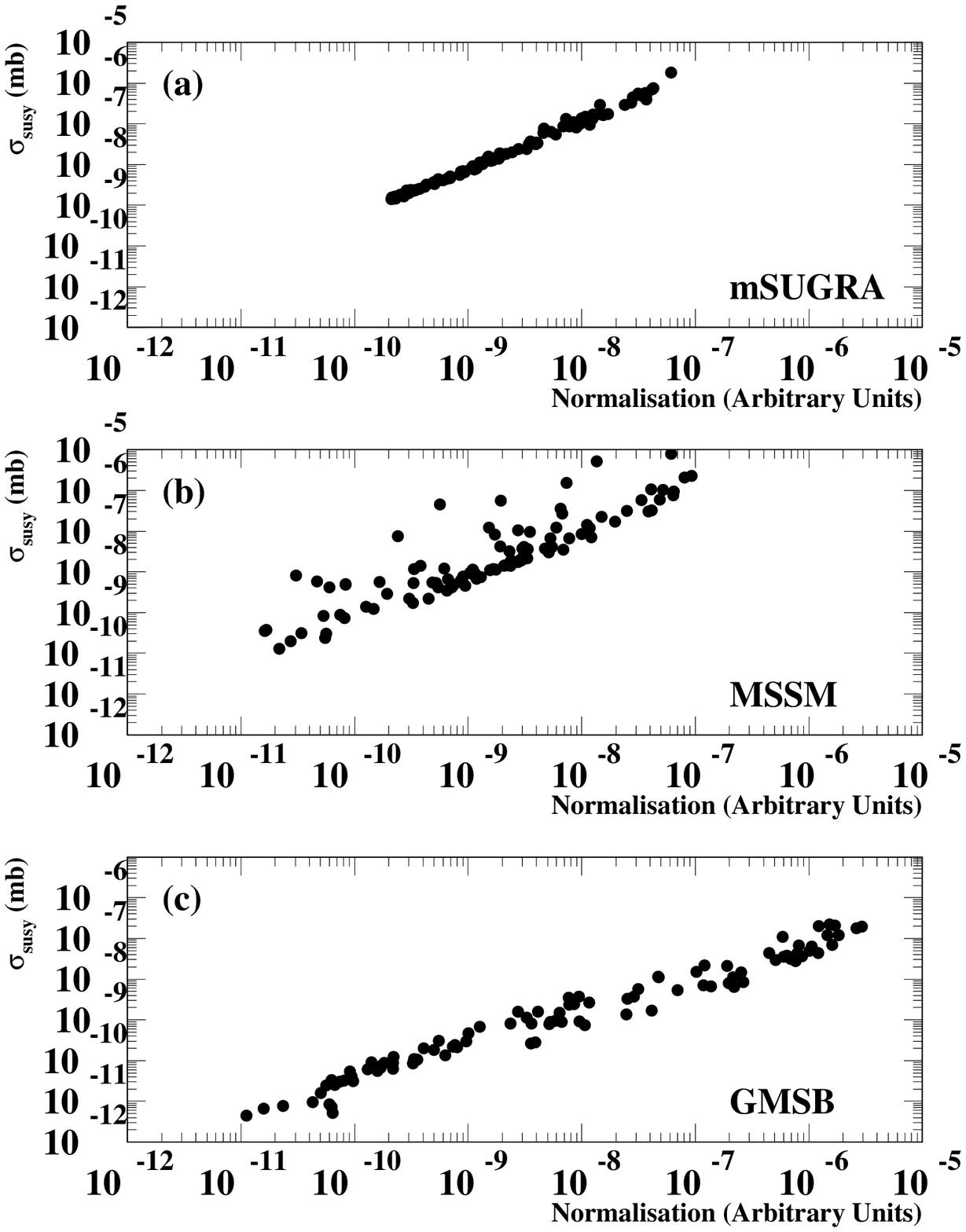,height=7.0in}
\caption{\label{fig4}}
\end{center}
\end{figure}

\begin{figure}
\begin{center}
\epsfig{file=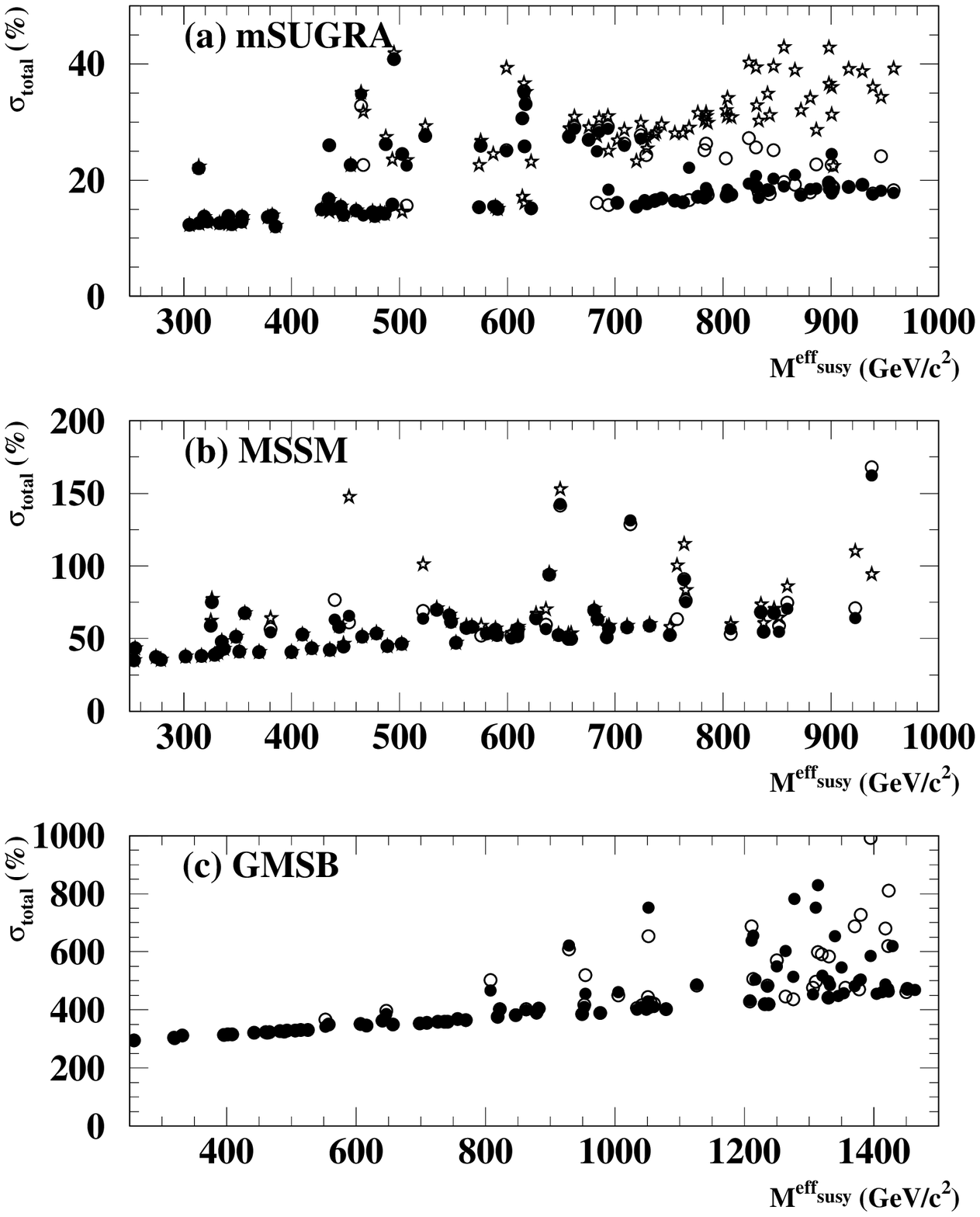,height=7.0in}
\caption{\label{fig5}}
\end{center}
\end{figure}

The above technique can also be used to measure the total SUSY
production cross section $\sigsus$, although in this case it is the
fitted normalisation of the signal $\mest$ distribution which is of
interest. The correlation between this normalisation and $\sigsus$ for
$\mest$ variable (3) is shown in Fig.~\ref{fig4}. The correlation is
reasonably good with the data best fitted by a power-law. For these
measurements the errors are non-gaussian (due to the power-law
relation between the normalisation and $\sigsus$) and in reality it is
the logarithm of the measured cross sections which is approximately
gaussian distributed. For this reason the intrinsic measurement
precisions for $\ln(\sigsus)$ are listed in Table~\ref{tab2}. The
overall (non-gaussian) precisions for $\sigsus$
(i.e. d$\sigsus$/$\sigsus$ = d(ln($\sigsus$))) are plotted in
Fig.~\ref{fig5} in the same format as Fig.~\ref{fig3}. For mSUGRA
models the overall $\sigsus$ measurement precision obtainable for 1000
fb$^{-1}$ is $\lesssim$ 15 \%, while for MSSM models it is $\lesssim$
50 \%.

Measurements of $\sigsus$ carried out in this way are inherently
sensitive to the type of SUSY model, in contrast to the measurements
of $\MSS$. This is because in some models (e.g. GMSB) the SUSY
particle decay characteristics can be such that the probability for
signal events to pass the selection cuts is reduced significantly
relative to that for mSUGRA models. The analysis presented here is
intended to be model independent and so projects data onto a single
axis perpendicular to the trendline of the mSUGRA models. Consequently
in the GMSB case, where the trend is very different from that for
mSUGRA, the presented measurement precision is poor ($\gtrsim$ 300
\%). If it were known that GMSB models were correct then an axis
perpendicular to the GMSB trendline could be used to obtain much
greater measurement precision ($\ln(\sigsus)$ precision $<$ 2.5
\%). This highlights the fact that in reality measurements of
$\sigsus$, unlike measurements of $\MSS$, are dependent on the assumed
SUSY model.

\section{Conclusions}

Model independent techniques for measuring the effective mass scale of
SUSY particles at the LHC have been investigated. Overall measurement
precisions better than 15 \% (40 \%) should be possible for mSUGRA
(MSSM) models after only one year of running at low
luminosity. Measurements should also be possible for models with rapid
decays to gravitino LSPs, although with the requirement of either
significantly increased statistics or measurement variables using
photon or lepton $p_T$. The total SUSY production cross section should
be measureable in a similar way ultimately to $\sim$ 15 \% (50 \%) in
mSUGRA (MSSM) models, although not in a completely model independent
manner.
 
\section*{Acknowledgments}
The author wishes to thank Frank Paige and Craig Buttar for their
careful reading of this manuscript and many helpful comments and
suggestions. He also wishes to acknowledge PPARC for support under the
Post-Doctoral Fellowship program.

\newpage

\section*{Tables}

Table 1: Estimates of the intrinsic $\MSS$ measurement precision for
mSUGRA, MSSM and GMSB models for the five $\mest$ variables discussed
in the text. The third and fourth columns show the fitted mean and
width of the projected $\MSS$ - $\mest$ correlation histogram for each
model and variable, and the fifth column their ratio. The sixth column
shows the expected fractional width estimated from the rms error on
the fitted means of the signal distributions. The seventh column
contains the intrinsic measurement precision estimated by subtracting
in quadrature column six from column five.  
\\
\newline
Table 2: Estimates of the intrinsic $\ln(\sigsus)$ measurement
precision for mSUGRA, MSSM and GMSB models for $\mest$ variable (3)
(defined in the text). The third and fourth columns show the fitted
mean and width of the projected $\ln(\sigsus)$ - normalisation
correlation histogram for each model and variable, and the fifth
column their ratio. The sixth column shows the expected fractional
width estimated from the rms error on the fitted normalisations of the
signal distributions. The seventh column contains the intrinsic
measurement precision estimated by subtracting in quadrature column
six from column five.

\newpage

\section*{Figures}

Figure 1: The effective SUSY mass scale $\MSS$ plotted against $\mest$
for variable (3) (defined in the text) for 100 random mSUGRA
(Fig.~1(a)), MSSM (Fig.~1(b)) and GMSB (Fig.~1(c)) models. Note the
differing scale in Fig.~1(c) due to the larger spread in $\mest$
values generated for GMSB models. In Fig.~1(c) those GMSB models where
the gaussian fit to the signal $\mest$ distribution failed due to
insufficient acceptance are omitted.  
\\
\newline
Figure 2: Projections of the points in Fig.~1 onto an axis transverse
to the fitted trendline of mSUGRA data (Fig.~1(a)) for $\mest$
variable (3). Fig.~2(a) shows the distribution for mSUGRA points,
Fig.~2(b) the distribution for MSSM points with $\MSS$ = $\MS$,
Fig.~2(c) the distribution for MSSM points with $\MSS$ given by
Eqn.~(\ref{eq3}) and Fig.~2(d) the distribution for GMSB points. Bin
widths are equal in Fig.~2(b) and Fig.~2(c) to aid comparison. Bin
widths differ between other plots.
\\
\newline
Figure 3: Overall precision for measurement of $\MSS$ after delivery of
integrated luminosities of 10 fb$^{-1}$ (stars), 100 fb$^{-1}$ (open
circles) and 1000 fb$^{-1}$ (filled circles) for $\mest$ variable
(3). Precisions for mSUGRA points are plotted in Fig.~3(a), MSSM
points in Fig.~3(b) and GMSB points in Fig.~3(c). No data are shown
for GMSB points for 10 fb$^{-1}$ integrated luminosity due to the poor
statistical significance of signal events in this scenario. Note the
differing scale in Fig.~3(c) due to the larger spread in $\mest$
values generated for GMSB models.
\\
\newline
Figure 4: The total SUSY particle production cross section $\sigsus$
plotted against the fitted normalisation of the signal distribution
for variable (3) (defined in the text) for 100 random mSUGRA
(Fig.~4(a)), MSSM (Fig.~4(b)) and GMSB (Fig.~4(c)) models. In
Fig.~4(c) those GMSB models where the gaussian fit to the signal
$\mest$ distribution failed due to insufficient acceptance are
omitted.
\\
\newline 
Figure 5: Overall (non-gaussian) precision for measurement of
$\sigsus$ after delivery of integrated luminosities of 10 fb$^{-1}$
(stars), 100 fb$^{-1}$ (open circles) and 1000 fb$^{-1}$ (filled
circles) for $\mest$ variable (3). Precisions for mSUGRA points are
plotted in Fig.~5(a), MSSM points in Fig.~5(b) and GMSB points in
Fig.~5(c). No data are shown for GMSB points for 10 fb$^{-1}$
integrated luminosity due to the poor statistical significance of
signal events in this scenario. Note the differing scale in Fig.~5(c)
due to the larger spread in $\mest$ values generated for GMSB models.
\newline
 
\newpage
\begin{table}[thb]
\begin{center}
\begin{tabular}{||l|c|c|c|c|c|c||} \hline\hline
Model	&Variable	&$\bar{x}$	&$\sigma$
&$\sigma/\bar{x}$	&rms error ($\bar{x}$)	&Precision (\%) \\ \hline
mSUGRA	&1	&1.585   &0.049   &0.031   &0.011   &2.9  \\
	&2	&0.991   &0.039   &0.039   &0.010   &3.8  \\
	&3	&1.700   &0.043   &0.026   &0.015   &2.1  \\
	&4	&1.089   &0.030   &0.028   &0.011   &2.5  \\
	&5	&1.168   &0.029   &0.025   &0.013   &2.1  \\ \hline
MSSM	&1	&1.657   &0.386   &0.233   &0.031   &23.1 \\
	&2	&0.998   &0.214   &0.215   &0.042   &21.1 \\
	&3	&1.722   &0.227   &0.132   &0.031   &12.8 \\
	&4	&1.092   &0.143   &0.131   &0.029   &12.8 \\
	&5	&1.156   &0.176   &0.152   &0.034   &14.8 \\ \hline
GMSB	&1	&1.660   &0.149   &0.090   &0.037   &8.1  \\ 
	&2	&1.095   &0.085   &0.077   &0.040   &6.6  \\
	&3	&1.832   &0.176   &0.096   &0.034   &9.0  \\
	&4	&1.235   &0.091   &0.074   &0.041   &6.1  \\
	&5	&1.273   &0.109   &0.086   &0.034   &7.9  \\ \hline\hline
\end{tabular}	
\vspace{1.0 cm}	
\caption{\label{tab1}{\it }}
\end{center}
\end{table}
\newpage
\begin{table}[thb]
\begin{center}
\begin{tabular}{||l|c|c|c|c|c|c||} \hline\hline
Model	&Variable	&$\bar{x}$	&$\sigma$
&$\sigma/\bar{x}$	&rms error ($\bar{x}$)	&Precision (\%) \\ \hline
mSUGRA	&3	&0.855	&0.008	&0.009	&0.003	&0.8 \\
MSSM 	&3	&0.848	&0.023	&0.027	&0.004	&2.7 \\
GMSB	&3	&0.742	&0.141	&0.190	&0.006	&19.0 \\ \hline\hline
\end{tabular}
\vspace{1.0 cm}
\caption{\label{tab2}{\it }}
\end{center}
\end{table}
\newpage
\newpage
\newpage
\newpage
\newpage

\end{document}